\documentclass[aps,prd,eqsecnum,twocolumn,showpacs,amsmath]{revtex4}
\usepackage[dvips]{color,graphicx}
\usepackage{amsfonts,amssymb,theorem,soul,xcolor,ulem}
\newcommand{\D}{{\rm d}}

{\theorembodyfont{\upshape}
}
{\theorembodyfont{\upshape}
}
{\theorembodyfont{\upshape}
}
{\theorembodyfont{\upshape}
}
{\theorembodyfont{\upshape}
}
{\theorembodyfont{\upshape}
}

\newcommand{\dalm}{\kern1pt\vbox{\hrule height 0.9pt\hbox{\vrule width
0.9pt\hskip 2.5pt\vbox{\vskip 5.5pt}\hskip 3pt\vrule width 0.3pt}\hrule height
0.3pt}\kern1pt}

\begin{document}

\title{
Planar black holes and wormholes with a flat exterior
}

\author{Hideki Maeda${}^{a}$}
\email{h-maeda@hgu.jp}
\author{Cristi{\'a}n Mart\'{\i}nez${}^{b,c}$}
\email{cristian.martinez@uss.cl}


\affiliation{
${}^a$ Department of Electronics and Information Engineering, Hokkai-Gakuen University, Sapporo 062-8605, Japan}
\affiliation{ 
${}^b$ Centro de Estudios Cient\'{\i}ficos (CECs), Avenida Arturo Prat 514, Valdivia, Chile}
\affiliation{
${}^c$Facultad de Ingenier\'{\i}a, Universidad San Sebasti\'an, General Lagos 1163, Valdivia, Chile}

\date{\today}

\begin{abstract}
We present $n(\ge 4)$-dimensional planar black holes and wormholes with a {\it flat} exterior, which are originated by an exact solution in general relativity.
The nonvacuum regions of these objects are described by the extended dynamical region inside a nondegenerate Killing horizon of Gamboa's static plane symmetric solution with a perfect fluid obeying a linear equation of state $p=\chi\rho$ for $\chi\in[-1/3,0)$. 
The matter field inside the horizon is not a perfect fluid but an anisotropic fluid that may be interpreted as a {\it spacelike} (tachyonic) perfect fluid.
While it satisfies the null and strong energy conditions in the black hole case, it violates all the standard energy conditions in the wormhole case.
The metric on the horizon is not analytic but at least $C^{1,1}$ in the single-null coordinates in both cases, so it is regular and there is no lightlike massive thin shell on the horizon.
\end{abstract}



\maketitle


\section{Introduction}

The interior structure of realistic black holes in our Universe is a highly nontrivial problem.
Although a curvature singularity exists inside the Schwarzschild or Kerr black hole, it is natural to assume that classical general relativity is no longer valid and the quantum effects of gravity dominate near the singularity.
Although the formulation of full quantum gravity is currently incomplete, a variety of metrics describing a nonsingular black hole have been proposed in the hope that curvature singularities will be avoided even in the classical modified theory of gravity realized in its low-energy limit.

Broadly speaking, global structures of spherically symmetric nonsingular black holes can be classified into two types: the regular-center type and the black-bounce type.
The former replaces the central timelike singularity inside the Reissner-Nordstr\"om black hole with a regular center. (See Sec.~2.2 of Ref.~\cite{Maeda:2021jdc} for a review.)
However, since this type of black holes is inevitably accompanied by an inner Cauchy horizon, it may not have a stable interior due to the mass-inflation instability~\cite{Poisson:1989zz,Poisson:1990eh,Ori:1991zz}.
(See Refs.~\cite{Rubio:2021obb,Bonanno:2020fgp,DiFilippo:2022qkl,Carballo-Rubio:2022kad,Carballo-Rubio:2024dca} for recent studies of the inner-horizon instability of the regular-center type.)

On the other hand, the latter black bounce typically replaces the spacelike singularity located in the dynamical region inside the Schwarzschild black hole by a regular big bounce.
Such a black hole was first referred to as a {\it black universe} by Bronnikov, Melnikov, and Dehnen~\cite{Bronnikov:2006fu}.
However, a different name ``{black bounce}'' introduced later by Simpson and Visser~\cite{Simpson:2018tsi} is more popular today.
The black bounce need not be accompanied by an inner horizon, and then it is free from the mass-inflation instability.

In recent studies, a novel black hole configuration called {\it fake Schwarzschild black hole} has been proposed~\cite{Maeda:2024tpl}.
The Schwarzschild vacuum solution describes the exterior of the event horizon of a fake Schwarzschild black hole, and the interior is described by a solution to the Einstein equations with a matter field.
The event horizon of this black hole is not analytic but regular described by the metric which is at least $C^{1,1}$ in a regular coordinate system.
Therefore, massive thin shells are not required for this configuration.
By construction, a fake Schwarzschild black hole shares the same thermodynamic properties with a Schwarzschild black hole, and those two black holes cannot be distinguished by observations.
One of the authors presented the first examples of exact solutions describing a fake Schwarzschild black hole using the extended dynamical region inside a Killing horizon of static perfect-fluid solutions obeying a linear equation of state~\cite{Maeda:2022lsm,Maeda:2024tpl}.
The metric describing a fake Schwarzschild black hole has also been studied with a different matter field in Ref.~\cite{Ovalle:2024wtv}.

While fake Schwarzschild black holes in the previous studies are spherically symmetric, we present in this article an even more bizarre configuration of plane symmetric black holes in $n(\ge 4)$ dimensions, based on the results in Ref.~\cite{Maeda:2025bnb}.
The exterior region of this new black hole is neither Schwarzschild nor its topological generalization but Minkowski.
We will show that the metric describes a plane symmetric black hole or wormhole depending on the parameters.
Our construction should be compared with the one in Ref.~\cite{Barcelo:2025egc} that requires one or two timelike massive thin shells to achieve a static axisymmetric black-hole spacetime in four dimensions where a part of the region outside the horizon is Minkowski.
Unlike this, the whole region outside the horizon of our black hole is Minkowski, and our construction does not require massive thin shells.

Our conventions for curvature tensors are $[\nabla _\rho ,\nabla_\sigma]V^\mu ={{R}^\mu }_{\nu\rho\sigma}V^\nu$ and ${R}_{\mu \nu }={{R}^\rho }_{\mu \rho \nu }$, where Greek indices run over all spacetime indices.
The signature of the Minkowski spacetime is $(-,+,+,\cdots,+)$.
We adopt the units such that $c=1$ and $\kappa_n:=8\pi G_n$, where $G_n$ is the $n$-dimensional gravitational constant.

\section{Minkowski spacetime in the quasiglobal coordinates}

\begin{figure}[htbp]
\begin{center}
\includegraphics[width=0.7\linewidth]{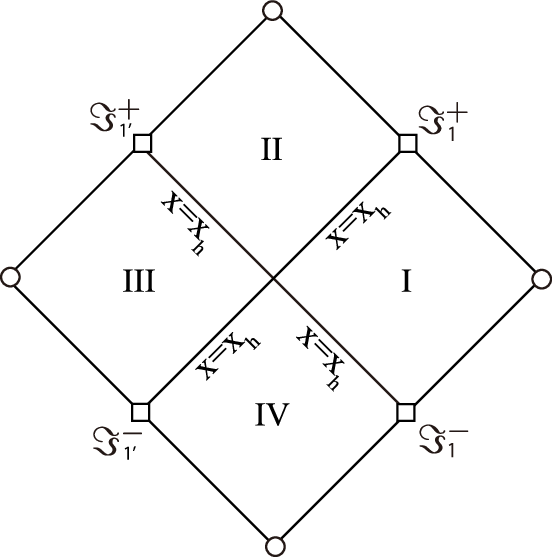}
\caption{
\label{Fig:PenroseDiagrams-Minkowski} The Penrose diagram of the Minkowski spacetime in the quasiglobal coordinates (\ref{metric-Buchdahl}).
$\Im^{+(-)}$ stands for a future (past) null infinity and its subscript is to distinguish different null infinities.}
\end{center}
\end{figure}

We begin by representing the $n$-dimensional Minkowski spacetime in the following quasiglobal coordinates with planar symmetry:
\begin{align}
&\D s^2=-H(x)\D t^2+H(x)^{-1}\D x^2+r(x)^2\D l_{n-2}^2, \label{metric-Buchdahl}
\end{align}
where $\D l_{n-2}^2$ is the line element on the $(n-2)$-dimensional flat space.
The Minkowski metric is given by
\begin{align}
&H(x)=H_1\Delta,\qquad r(x)=r_0,\label{flat}
\end{align}
where $\Delta:=x-x_{\rm h}$ and $r_0$, $H_1$, and $x_{\rm h}$ are arbitrary constants.
As shown below, $x=x_{\rm h}$ is the Rindler horizon and the region where $H_1(x-x_{\rm h})>(<)0$ holds corresponds to the regions I and III (II and IV) in the Penrose diagram drawn in Fig.~\ref{Fig:PenroseDiagrams-Minkowski}, which are referred to as the Rindler (Milne) charts.

By the following coordinate transformations:
\begin{align}
t=\frac{2}{|H_1|}{\bar t},\qquad H_1(x-x_{\rm h})=\frac{H_1^2}{4}{\bar x}^2,
\end{align}
the region where $H_1(x-x_{\rm h})>0$ holds is represented in the Rindler coordinates as
\begin{align}
&\D s^2=-{\bar x}^2\D {\bar t}^2+\D {\bar x}^2+r_0^2\D l_{n-2}^2.
\end{align}
By another set of coordinate transformations,
\begin{align}
{\bar t}=\mbox{arctanh}\biggl(\frac{T}{X}\biggl),\qquad {\bar x}=\sqrt{X^2-T^2},
\end{align}
we obtain
\begin{align}
&\D s^2=-\D T^2+\D X^2+r_0^2\D l_{n-2}^2. \label{flat-Cartesian}
\end{align}
The Rindler coordinates cover the domain given by $X^2>T^2$ and $x=x_{\rm h}$ corresponds to the Rindler horizon $X=\pm T$.

By the following coordinate transformations:
\begin{align}
t=\frac{2}{|H_1|}{\hat x},\qquad H_1(x-x_{\rm h})=-\frac{H_1^2}{4}{\hat t}^2,
\end{align}
the region where $H_1(x-x_{\rm h})<0$ holds is represented in the Milne coordinates as
\begin{align}
&\D s^2=-\D {\hat t}^2+{\hat t}^2\D {\hat x}^2+r_0^2\D l_{n-2}^2.
\end{align}
By another set of coordinate transformations,
\begin{align}
{\hat t}=\sqrt{T^2-X^2},\qquad {\hat x}=\mbox{arctanh}\biggl(\frac{X}{T}\biggl),
\end{align}
we obtain the metric~(\ref{flat-Cartesian}).
The Milne coordinates cover the domain given by $T^2>X^2$.

In the quasiglobal coordinates (\ref{metric-Buchdahl}), $x=x_{\rm h}$ determined by $H(x_{\rm h})=0$ is a coordinate singularity.
This coordinate singularity can be removed by introducing an ingoing null coordinate $v:=t\pm \int H(x)^{-1}\D x$, with which the metric is given as
\begin{equation}
\D s^2=-H(x)\D v^2\pm2\D v\D x+r(x)^2\D l_{n-2}^2.\label{metric-Buchdahl-v}
\end{equation}
If $x=x_{\rm h}$ is regular and $r(x_{\rm h})\ne 0$ holds in the single-null coordinates (\ref{metric-Buchdahl-v}), it is a Killing horizon associated with a Killing vector $\xi^\mu\partial_\mu=\partial_v$ which is a null hypersurface.
As shown in Fig.~\ref{Fig:PenroseDiagrams-Minkowski}, the Rindler horizon is a Killing horizon but not an event horizon.
In particular, the points where a square is placed in Fig.~\ref{Fig:PenroseDiagrams-Minkowski} are neither timelike infinities in the regions I and III nor spacelike infinities in the regions II and IV, but just parts of the null infinities.
The surface gravity $\kappa$ on the Killing horizon is computed to give 
\begin{equation}
\kappa=\pm \frac12H'|_{x=x_{\rm h}}=\pm\frac12H_1\label{kappa-flat}
\end{equation}
from the definition $\xi^\nu\nabla_\nu\xi^\mu|_{x=x_{\rm h}}=\kappa\xi^\mu|_{x=x_{\rm h}}$.

\section{Dynamical region of the Gamboa solution}
In this work, we study an exact static plane symmetric solution to the $n(\ge 4)$-dimensional Einstein equations with a perfect fluid obeying $p=\chi\rho$ given by 
\begin{align}
\label{EFE-0}
\begin{aligned}
&G_{\mu\nu}=\kappa_nT_{\mu\nu},\\
&{T}_{\mu\nu}=(\rho+p) u_\mu u_\nu+pg_{\mu\nu}.
\end{aligned}
\end{align}
Here $\rho$ and $p$ are the energy density and pressure of a perfect fluid, respectively, and $u^\mu$ is the normalized $n$-velocity of the fluid element satisfying $u_\mu u^\mu=-1$.

As shown in Appendix~B in Ref.~\cite{Maeda:2024lbq}, there is no static plane symmetric solution for $\chi=0$.
Then, by Proposition~2 in Ref.~\cite{Maeda:2024lbq}, any static plane symmetric solution in this system does not admit a Killing horizon unless $-1/3\le \chi< 0$ or $\chi=-1$.
Since a perfect fluid for $\chi=-1$ is equivalent to a cosmological constant $\Lambda$, we will assume $\chi\in[-1/3,0)$ hereafter.

In particular, we study the Gamboa solution~\cite{GamboaSaravi} for $\chi\in[-1/3,0)$ in the following form~\cite{Maeda:2025bnb}:
\begin{align}
\label{ansatz00-2}
\begin{aligned} 
&\D s^2=-A\D T^2+B\D r^2+r^2\D l_{n-2}^2,\\
&A(r)=\mp\frac{2M}{r^{n-3}}(\pm \Pi)^{1/(2+\beta)},\\
&B(r)=\mp\frac{r^{n-3}}{2M}(\pm \Pi)^{-(3+2\beta)/(2+\beta)},\\
&\rho=\frac{p}{\chi}=\frac{\alpha M}{r^{(n-3)(1+\beta)}}(\pm \Pi)^{(1+\beta)/(2+\beta)},
\end{aligned} 
\end{align} 
where the upper (lower) sign is taken for $\chi\in[-1/3,\chi_0)$ and $\chi\in[\chi_0,0)$, respectively, and $\Pi(r)$ is defined by 
\begin{align}
\label{def-Pi}
&\Pi(r):=\left\{
\begin{array}{ll}
1-h_1/r^{(n-3)\beta-2} & [\chi\ne \chi_0]\\
1-h_1\ln|r| & [\chi=\chi_0]
\end{array}
\right..
\end{align} 
Here $\chi_0$, $\beta$, and $\alpha$ are defined by 
\begin{align} 
&\chi_0:=-\frac{n-3}{3n-5}\left(>-\frac13\right),\label{def-chi0}\\
&\beta:=-\frac{1+3\chi}{2\chi}~~\Leftrightarrow~~\chi=-\frac{1}{3+2\beta},\label{def-beta}\\
&\alpha=\left\{
\begin{array}{ll}
-(n-2) \zeta h_1/[\kappa_n \chi (1-\chi)] & [\chi\ne \chi_0]\\
-(3n-5)h_1/(2\kappa_n) & [\chi=\chi_0]
\end{array}
\right.,\label{def-alpha}
\end{align} 
where
\begin{align}
\zeta:=(3n-5)\chi+(n-3)=\frac{2(n-3)\beta-4}{3+2\beta}.\label{def-zeta}
\end{align} 
We note that $\beta\in[0,\infty)$ and $\beta=2/(n-3)$ correspond to $\chi\in[-1/3,0)$ and $\chi=\chi_0$, respectively.
\begin{table*}[htb]
\begin{center}
\caption{Domains of $r$ of the Gamboa solution where $A<0$ and $B<0$ hold and their corresponding domains in the quasiglobal coordinates (\ref{metric-Buchdahl}).}
\label{table:correspondence1}
\begin{tabular}{|c|c|c|c|c|c|c|c|}\hline
$\chi$ & $M$ & $\Pi_1$ & $\D r/\D x$ & Sign in Eq.~(\ref{ansatz00-2}) & $\rho$ & Domain of $r$ & Domain of $x$ \\ \hline
$[-1/3,\chi_0)$ & $+$ & $-$ & $+$ & upper & $-$ & $r<r_{\rm h}$ & $x<x_{\rm h}$ \\ \cline{2-8}
& $-$ & $-$ & $-$ & lower & $+$ & $r>r_{\rm h}$ & $x<x_{\rm h}$ \\ \cline{1-8}
$\chi_0$ with $h_1>0$ & $+$ & $-$ & $+$ & upper & $-$ & $r<r_{\rm h}$ & $x<x_{\rm h}$ \\ \cline{2-8}
& $-$ & $-$ & $-$ & lower & $+$ & $r>r_{\rm h}$ & $x<x_{\rm h}$ \\ \cline{1-8}
$\chi_0$ with $h_1<0$ & $-$ & $+$ & $+$ & lower & $-$ & $r<r_{\rm h}$ & $x<x_{\rm h}$ \\ \cline{2-8}
& $+$ & $+$ & $-$ & upper & $+$ & $r>r_{\rm h}$ & $x<x_{\rm h}$ \\ \cline{1-8}
$[\chi_0,0)$ & $-$ & $+$ & $+$ & lower & $-$ & $r<r_{\rm h}$ & $x<x_{\rm h}$ \\ \cline{2-8}
& $+$ & $+$ & $-$ & upper & $+$ & $r>r_{\rm h}$ & $x<x_{\rm h}$ \\ \hline
\end{tabular}
\end{center}
\end{table*}

With $h_1=0$, the Gamboa solution (\ref{ansatz00-2}) with the upper sign is identical to the topological Schwarzschild-Tangherlini vacuum solution.
In this work, we consider domains where $A<0$ and $B<0$ hold, and hence the spacetime is dynamical as $T$ and $r$ are spacelike and timelike coordinates, respectively.
As a consequence, the matter field in this dynamical region is not a perfect fluid but an anisotropic fluid with the energy density $\mu$, radial pressure $p_1$, and tangential pressure $p_2$ given by
\begin{align}
\label{matter-interior}
&\mu=-p, \qquad p_1=-\rho,\qquad p_2=p,
\end{align} 
where $p$ and $\rho$ are given by Eq.~(\ref{ansatz00-2})~\cite{Maeda:2024tpl}.
Alternatively, the matter field in the dynamical region may be interpreted as a {\it spacelike} (or {\it tachyonic}) perfect fluid~\cite{Maeda:2024tpl}.
The standard energy conditions consist of the {\it null} energy condition (NEC), {\it weak} energy condition (WEC), {\it dominant} energy condition (DEC), and {\it strong} energy condition (SEC)~\cite{Maeda:2018hqu}.
For the matter field in the dynamical region, by Eq.~(\ref{matter-interior}) and the result in Sec.~3.1 in Ref.~\cite{Maeda:2018hqu}, equivalent representations of those energy conditions are given by
\begin{align}
\mbox{NEC}:&~~(1+\chi)\rho\le 0,\label{NEC-I-interior}\\
\mbox{WEC}:&~~\chi \rho\le 0\mbox{~in addition to NEC},\label{WEC-I-interior}\\
\mbox{DEC}:&~~(1-\chi)\rho\ge 0\mbox{~in addition to WEC},\label{DEC-I-interior}\\
\mbox{SEC}:&~~(1-\chi)\rho\le 0\mbox{~in addition to NEC}.\label{SEC-I-interior}
\end{align}

We define $r_{\rm h}$ by $A(r_{\rm h})=0=\Pi(r_{\rm h})$ and hence
\begin{align}
\label{def-rh}
&r_{\rm h}:=\left\{
\begin{array}{ll}
h_1^{1/[(n-3)\beta-2]} & [\chi\ne \chi_0]\\
e^{1/h_1} & [\chi=\chi_0]
\end{array}
\right..
\end{align} 
In order to have a real positive $r_{\rm h}$, we assume $h_1>0$ for $\chi\ne \chi_0$ and $h_1\ne 0$ for $\chi=\chi_0$.
Near $r=r_{\rm h}$, we obtain 
\begin{align}
\Pi\simeq \Pi_1(r-r_{\rm h}),
\end{align} 
where
\begin{align}
\label{def-Pi1}
&\Pi_1:=\left\{
\begin{array}{ll}
[(n-3)\beta-2]/r_{\rm h} & [\chi\ne \chi_0]\\
-h_1/r_{\rm h} & [\chi=\chi_0]
\end{array}
\right..
\end{align} 
The conditions $A<0$ and $B<0$ are satisfied in the domain $r<r_{\rm h}$ only in the following four cases.
\begin{itemize}
\item For $\chi\in[-1/3,\chi_0)$ with $M>0$ and the upper sign of Eq.~(\ref{ansatz00-2}).

\item For $\chi=\chi_0$ with $h_1>0$ and $M>0$ and the upper sign of Eq.~(\ref{ansatz00-2}).

\item For $\chi=\chi_0$ with $h_1<0$ and $M<0$ and the lower sign of Eq.~(\ref{ansatz00-2}).

\item For $\chi\in (\chi_0,0)$ with $M<0$ and the lower sign of Eq.~(\ref{ansatz00-2}).
\end{itemize}
In this domain, as $\rho<0$ is satisfied, the NEC and SEC are satisfied, while the WEC and DEC are violated.

The conditions $A<0$ and $B<0$ are satisfied in the domain $r>r_{\rm h}$ only in the following four cases.
\begin{itemize}
\item For $\chi\in[-1/3,\chi_0)$ with $M<0$ and the lower sign of Eq.~(\ref{ansatz00-2}).

\item For $\chi=\chi_0$ with $h_1>0$ and $M<0$ and the lower sign of Eq.~(\ref{ansatz00-2}).

\item For $\chi=\chi_0$ with $h_1<0$ and $M>0$ and the upper sign of Eq.~(\ref{ansatz00-2}).

\item For $\chi\in (\chi_0,0)$ with $M>0$ and the upper sign of Eq.~(\ref{ansatz00-2}).
\end{itemize}
In this domain, as $\rho>0$ is satisfied, all the standard energy conditions are violated.
The results are summarized in Table~\ref{table:correspondence1}.

By the following coordinate transformations and an identification:
\begin{align}
&T=\varepsilon \Omega t,\label{trans-t}\\
&x=\Omega\int (\pm \Pi)^{-(1+\beta)/(2+\beta)}\D r, \label{trans1}\\
&H(x)\equiv \Omega^2 A(r(x)), \label{trans2}
\end{align} 
the Gamboa solution (\ref{ansatz00-2}) is mapped into the quasiglobal coordinates (\ref{metric-Buchdahl}), where $r=r_{\rm h}$ corresponds to $x=x_{\rm h}$ satisfying $H(x_{\rm h})=0$.
Here $\Omega$ is a gauge constant and $\varepsilon =\pm 1$ is chosen such that $\D T/\D t>1$ holds.
By Proposition~6 in Ref.~\cite{Maeda:2024lbq}, $r=r_{\rm h}$ is a nondegenerate Killing horizon in the regular coordinate system (\ref{metric-Buchdahl-v}) for $\chi\in[-1/3,0)$, or equivalently $\beta\in[0,\infty)$.

With $\Omega=-(2+\beta)/(M\Pi_1)$, Eq.~(\ref{trans1}) gives
\begin{align}
&x-x_{\rm h}\simeq \mp \frac{(2+\beta)^2}{M\Pi_1^2}[\pm \Pi_1(r-r_{\rm h})]^{1/(2+\beta)}\label{x-r1}
\end{align} 
near $x=x_{\rm h}$.
Then, Eqs.~(\ref{trans-t})--(\ref{trans2}) give the asymptotic expansion of the solution near $x=x_{\rm h}$ as
\begin{align}
\label{asymp-H1}
\begin{aligned}
&r\simeq r_{\rm h}\pm\frac{(\mp M\Pi_1^2\Delta)^{2+\beta}}{\Pi_1(2+\beta)^{2(2+\beta)}},\\
&H\simeq\frac{2}{r_{\rm h}^{n-3}}\Delta+O(\Delta^{3+\beta}),\\
&\rho\simeq \frac{\alpha M(\mp M\Pi_1^2\Delta)^{1+\beta}}{(2+\beta)^{2(1+\beta)}r_{\rm h}^{(n-3)(1+\beta)}}.
\end{aligned} 
\end{align} 
By Eq.~(\ref{x-r1}), the Gamboa solution~(\ref{ansatz00-2}) with the parameters under consideration summarized in Table~\ref{table:correspondence1} is mapped into the domain $x<x_{\rm h}$.
Then, by Eq.~(\ref{trans1}), $r(x)$ is a monotonically increasing (decreasing) function for $M\Pi_1<(>)0$ as $\D r/\D x>(<)0$ holds.

In the Gamboa solution, $r=0$ corresponds to a curvature singularity since $\rho$ blows up there.
As shown in Sec.~IID in~Ref.~\cite{Maeda:2025bnb} for $\chi\in[-1/3,0)$, $r\to \infty$ is null infinity and causally null in the Penrose diagram for $\chi\in[-1/3,0)$.
On the other hand, $r=0$ is not null infinity and causally non-null in the Penrose diagram for $\chi\in[-1/3,0)$.

\section{Planar black holes and wormholes with flat exterior}

Now, using the Gamboa solution for $\chi\in[-1/3,0)$, we construct plane symmetric spacetimes describing a black hole or wormhole with the flat exterior.
In general, a closed-form expression of the integral~(\ref{trans1}) is not available.
However, it is available with $\beta=4/(n-4)$, or equivalently $\chi=-(n-4)/(3n-4)(>\chi_0)$, for $n\ge 5$.
We first discuss this special case and then deal with the most general case.

\subsection{For $\chi=-(n-4)/(3n-4)$ with $n\ge 5$}
\label{sec:example-exact}

In this special case, by coordinate transformations given by 
\begin{align}
\label{x-r-exact}
\begin{aligned}
&t=- \varepsilon MT,\\
&x=x_{\rm h}\mp M^{-1}\left[\pm(r^{2+\beta}-h_1)\right]^{1/(2+\beta)},
\end{aligned} 
\end{align} 
the Gamboa solution~(\ref{ansatz00-2}) is written in the quasiglobal coordinates (\ref{metric-Buchdahl}) as
\begin{align}
\label{Gamboa-H-interior}
\begin{aligned}
&r(x)=\left[h_1-{\bar m}(-\Delta)^{2+\beta}\right]^{1/(2+\beta)},\\
&H(x)=\frac{2\Delta}{r^{2(2+\beta)/\beta}}\biggl(=\frac{2\Delta}{r^{n-2}}\biggl),\\
&\rho(x)=\frac{p}{\chi}=-\frac{\alpha{\bar m}(-\Delta)^{1+\beta}}{r^{2(1+\beta)(2+\beta)/\beta}},
\end{aligned} 
\end{align} 
where $\Delta:=x-x_{\rm h}$ and ${\bar m}:=\mp(\pm M)^{2+\beta}$.
Here $x_{\rm h}$ is an integration constant of the integral~(\ref{trans1}), so that $r=r_{\rm h}(=h_1^{1/(2+\beta)})$ corresponds to $x=x_{\rm h}$.
This form of the solution is parametrized by two real constants $h_1$ and ${\bar m}$.

We note that the solution (\ref{Gamboa-H-interior}) with $h_1>0$ and ${\bar m}=0$ is locally Minkowski although it is not realized in the parameter space of the Gamboa solution in the original form (\ref{ansatz00-2}).
Now suppose that the parameters in the domains $x> x_{\rm h}$ and $x< x_{\rm h}$ are given by $(h_1,m)=(h_1^+,0)$ and $(h_1,m)=(h_1^-,m_-)$, respectively, where $h_1^+=h_1^-(>0)$ and $m_-\ne 0$ are satisfied.
Then, the horizon $x=x_{\rm h}$ is regular because the metric in the single-null coordinates (\ref{metric-Buchdahl-v}) is at least $C^{1,1}$ there.
The resulting metric describes a black hole or a wormhole with the Minkowski (flat) exterior, where a nondegenerate Killing horizon $x=x_{\rm h}$ coincides with the event horizon in the black hole case.

\begin{figure}[htbp]
\includegraphics[width=0.35\textwidth]{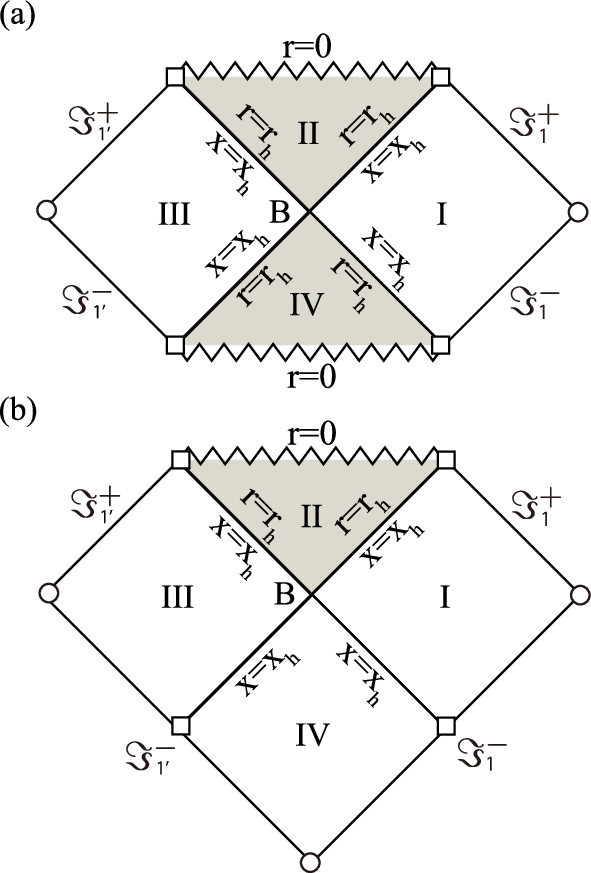}
\caption{
\label{Fig:PenroseDiagrams-FakeMinkowskiBH} The Penrose diagram of a planar black hole with the Minkowski exterior (a) in the regions I and III and (b) in the regions I, III, and IV.
The shaded regions are described by the Gamboa solution (\ref{ansatz00-2}).
A zigzag line corresponds to a curvature singularity located at $r=0$.}
\end{figure}

In particular, the spacetime describes a black hole for ${\bar m}>0$, which is realized for $M<0$ with the lower sign in the solution~(\ref{ansatz00-2}).
In this case, $r(x)$ increases monotonically in the domain $x<x_{\rm h}$ and there appears a spacelike curvature singularity at $x=x_{\rm s}$ determined by $r(x_{\rm s})=0$.
The Penrose diagram of this planar black hole is drawn in Fig.~\ref{Fig:PenroseDiagrams-FakeMinkowskiBH}.
Inside the horizon, while the NEC and SEC are satisfied, the WEC and DEC are violated.

\begin{figure}[htbp]
\includegraphics[width=0.35\textwidth]{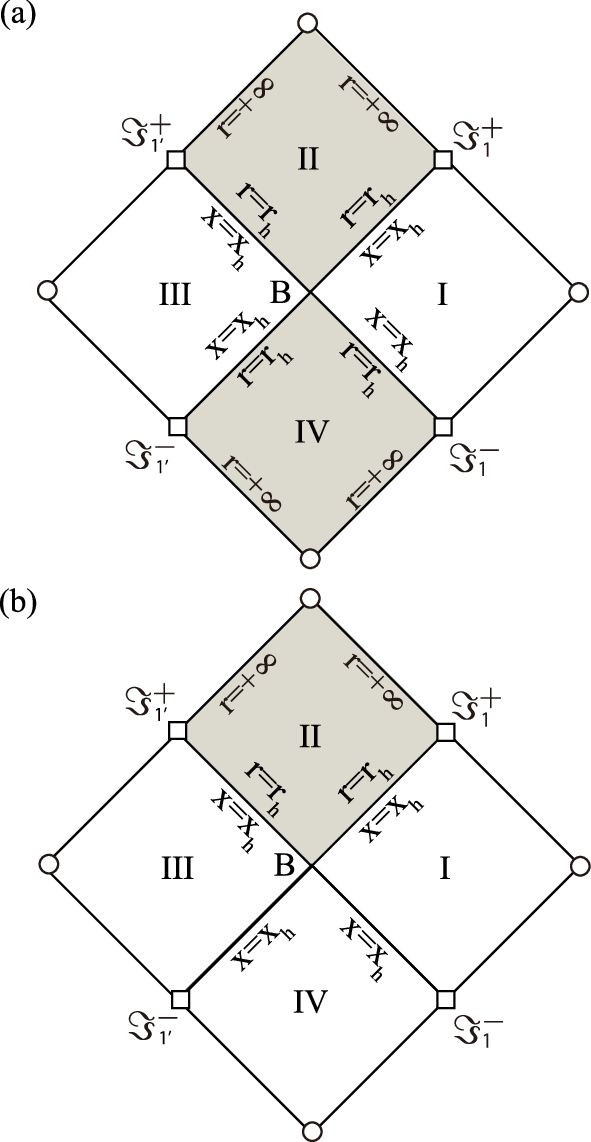}
\caption{
\label{Fig:PenroseDiagrams-FakeMinkowskiBB} The Penrose diagram of a planar wormhole with the Minkowski exterior (a) in the regions I and III and (b) in the regions I, III, and IV.
The shaded regions are described by the Gamboa solution (\ref{ansatz00-2}).
The Killing horizon $x=x_{\rm h}$ is not an event horizon in these cases. }
\end{figure}

In contrast, the spacetime describes a globally regular wormhole for ${\bar m}<0$, which is realized for $M>0$ with the upper sign in the solution~(\ref{ansatz00-2}).
In this case, $r(x)$ decreases monotonically in the domain $x<x_{\rm h}$.
The Penrose diagram of this planar wormhole is drawn in Fig.~\ref{Fig:PenroseDiagrams-FakeMinkowskiBB}.
The Killing horizon $x=x_{\rm h}$ is not an event horizon because the points where a square is placed in Fig.~\ref{Fig:PenroseDiagrams-FakeMinkowskiBB} are neither timelike infinities in the regions I and III nor spacelike infinities in the regions II and IV, but just parts of the null infinities $\Im^{+(-)}_1$ and $\Im^{+(-)}_{1'}$.
All the energy conditions are violated inside the horizon in this case.

\subsection{General $\chi\in[-1/3,0)$}

In the most general case $\chi\in[-1/3,0)$, suppose that the domains $x<x_{\rm h}$ and $x>x_{\rm h}$ in the single-null coordinates (\ref{metric-Buchdahl-v}) are described by the Gamboa solution (\ref{ansatz00-2}) and the Minkowski solution (\ref{metric-Buchdahl}), respectively.
Since $x_{\rm h}$, $H_1$, and $r_0$ are arbitrary in the Minkowski solution, we can set $H_1$ and $r_0$ such that $H_1=2/r_{\rm h}^{n-3}$ and $r_0=r_{\rm h}$.
Then, the asymptotic expansion (\ref{asymp-H1}) near $x=x_{\rm h}$ shows that the metric (\ref{metric-Buchdahl-v}) in the single-null coordinates is at least $C^{1,1}$ at $x=x_{\rm h}$, so that it is a regular nondegenerate Killing horizon.

With the parameters and sign of the solution (\ref{ansatz00-2}) that correspond to the domain $r<r_{\rm h}$ in Table~\ref{table:correspondence1}, the resulting spacetime describes a planar black hole with the Minkowski (flat) exterior and a spacelike curvature singularity inside the horizon, of which Penrose diagram is drawn in Fig.~\ref{Fig:PenroseDiagrams-FakeMinkowskiBH}.
While the NEC and SEC are satisfied in the interior dynamical region, the WEC and DEC are violated inside the horizon.

In contrast, with the choice of the parameters and sign of the solution (\ref{ansatz00-2}) that correspond to the domain $r>r_{\rm h}$ in Table~\ref{table:correspondence1}, the resulting spacetime describes a globally regular planar wormhole, of which Penrose diagram is drawn in Fig.~\ref{Fig:PenroseDiagrams-FakeMinkowskiBB}.
All the standard energy conditions are violated in the interior dynamical region.

It is emphasized that the regions II and IV in Fig.~\ref{Fig:PenroseDiagrams-FakeMinkowskiBH}(a) and Fig.~\ref{Fig:PenroseDiagrams-FakeMinkowskiBB}(a) may be described by the Gamboa solutions for different values of $\chi\in[-1/3,0)$.
Also, looking at Fig.~\ref{Fig:PenroseDiagrams-FakeMinkowskiBH}(b) or Fig.~\ref{Fig:PenroseDiagrams-FakeMinkowskiBB}(b), one might expect that the bifurcation $(n-2)$-surface, denoted by B, is singular.
Nevertheless, it is regular contrary to intuition~\cite{Maeda:2025bnb}.
A physical explanation of this counterintuitive regularity was provided in Ref.~\cite{Maeda:2024tpl} based on the fact that the matter field in the dynamical region can be interpreted as a spacelike (tachyonic) perfect fluid.
Such a fluid element in the region II moves in a spacelike direction and does not cross a Killing horizon and a bifurcation surface.
Since orbits of the fluid elements do not have a past endpoint at B, the bifurcation $(n-2)$-surface B can be regular.

\section{Summary and open questions}

In this article, using the extended dynamical region inside a nondegenerate Killing horizon of Gamboa's static plane symmetric solution with a perfect fluid obeying a linear equation of state $p=\chi\rho$ for $\chi\in[-1/3,0)$, we have constructed $n(\ge 4)$-dimensional planar black holes and wormholes with the Minkowski exterior in general relativity.
In this construction, the matter field, which may be interpreted as a spacelike (tachyonic) perfect fluid, violates the WEC and DEC in the black hole case and all the standard energy conditions in the wormhole case.

In fact, this configuration of black holes and wormholes are not possible under the DEC by the conservation theorem on page 94 in the textbook~\cite{Hawking:1973uf} showing that, if $T_{\mu\nu}=0$ holds on a closed achronal hypersurface $\Sigma$, $T_{\mu\nu}=0$ holds everywhere in the domain of dependence of $\Sigma$. (See Proposition~2.2.3 in Ref.~\cite{Iizuka:2025xnd}.)
Since we can prepare a closed achronal hypersurface $\Sigma$ with $T_{\mu\nu}=0$ that spans the regions I, IV, and III and contains the region IV in its domain of dependence in Fig.~\ref{Fig:PenroseDiagrams-FakeMinkowskiBH}(b) and Fig.~\ref{Fig:PenroseDiagrams-FakeMinkowskiBB}(b), the DEC is violated at least somewhere in the region IV by the contraposition of Proposition~2.2.3 in Ref.~\cite{Iizuka:2025xnd}.

An open issue is the thermodynamics of the black hole presented in this work. 
Although we have considered only the framework of general relativity in $n \ge 4$ dimensions,  such an analysis could be nontrivial.  
We leave a detailed study of thermodynamical aspects of our black hole for future investigations.

\acknowledgments

H.M. thanks Tomohiro Harada for discussions and is very grateful to Centro de Estudios Cient\'{\i}ficos for hospitality and support, where a large part of this work was carried out. This work has been partially supported by the ANID FONDECYT grants 1220862 and 1241835.




\begin{references}



\bibitem{Maeda:2021jdc}
H.~Maeda,
``Quest for realistic nonsingular black-hole geometries: regular-center type'',
JHEP \textbf{11}, 108 (2022)
doi:10.1007/JHEP11(2022)108
[arXiv:2107.04791 [gr-qc]].


\bibitem{Poisson:1989zz}
E.~Poisson and W.~Israel,
``Inner-horizon instability and mass inflation in black holes'',
Phys. Rev. Lett. \textbf{63}, 1663-1666 (1989)
doi:10.1103/PhysRevLett.63.1663

\bibitem{Poisson:1990eh}
E.~Poisson and W.~Israel,
``Internal structure of black holes'',
Phys. Rev. D \textbf{41}, 1796-1809 (1990)
doi:10.1103/PhysRevD.41.1796


\bibitem{Ori:1991zz}
A.~Ori,
``Inner structure of a charged black hole: An exact mass-inflation solution'',
Phys. Rev. Lett. \textbf{67}, 789-792 (1991)
doi:10.1103/PhysRevLett.67.789


\bibitem{Bonanno:2020fgp}
A.~Bonanno, A.~P.~Khosravi and F.~Saueressig,
``Regular black holes with stable cores'',
Phys. Rev. D \textbf{103}, no.12, 124027 (2021)
doi:10.1103/PhysRevD.103.124027
[arXiv:2010.04226 [gr-qc]].


\bibitem{Rubio:2021obb}
R.~Carballo-Rubio, F.~Di Filippo, S.~Liberati, C.~Pacilio and M.~Visser,
``Inner horizon instability and the unstable cores of regular black holes'',
JHEP \textbf{05}, 132 (2021)
doi:10.1007/JHEP05(2021)132
[arXiv:2101.05006 [gr-qc]].

\bibitem{DiFilippo:2022qkl}
F.~Di Filippo, R.~Carballo-Rubio, S.~Liberati, C.~Pacilio and M.~Visser,
``On the Inner Horizon Instability of Non-Singular Black Holes'',
Universe \textbf{8} (2022) no.4, 204
doi:10.3390/universe8040204
[arXiv:2203.14516 [gr-qc]].


\bibitem{Carballo-Rubio:2022kad}
R.~Carballo-Rubio, F.~Di Filippo, S.~Liberati, C.~Pacilio and M.~Visser,
``Regular black holes without mass inflation instability'',
JHEP \textbf{09} (2022), 118
doi:10.1007/JHEP09(2022)118
[arXiv:2205.13556 [gr-qc]].

\bibitem{Carballo-Rubio:2024dca}
R.~Carballo-Rubio, F.~Di Filippo, S.~Liberati and M.~Visser,
``Mass Inflation without Cauchy Horizons'',
Phys. Rev. Lett. \textbf{133}, no.18, 181402 (2024)
doi:10.1103/PhysRevLett.133.181402
[arXiv:2402.14913 [gr-qc]].


\bibitem{Bronnikov:2006fu}
K.~A.~Bronnikov, V.~N.~Melnikov and H.~Dehnen,
``Regular black holes and black universes'',
Gen. Rel. Grav. \textbf{39} (2007), 973-987
doi:10.1007/s10714-007-0430-6
[arXiv:gr-qc/0611022 [gr-qc]].



\bibitem{Simpson:2018tsi}
A.~Simpson and M.~Visser,
``Black-bounce to traversable wormhole'',
JCAP \textbf{02} (2019), 042
doi:10.1088/1475-7516/2019/02/042
[arXiv:1812.07114 [gr-qc]].


\bibitem{Maeda:2024tpl}
H.~Maeda,
``Fake Schwarzschild and Kerr black holes'',
[arXiv:2410.11937 [gr-qc]].


\bibitem{Maeda:2022lsm}
H.~Maeda,
``Vacuum-dual static perfect fluid obeying $p=-(n-3)\rho/(n+1)$ in $n(\ge 4)$ dimensions'',
Class. Quant. Grav. \textbf{40}, no.8, 085014 (2023)
doi:10.1088/1361-6382/acc3f1
[arXiv:2210.10795 [gr-qc]].



\bibitem{Ovalle:2024wtv}
J.~Ovalle,
``Schwarzschild black hole revisited: Before the complete collapse'',
Phys. Rev. D \textbf{109}, no.10, 104032 (2024)
doi:10.1103/PhysRevD.109.104032
[arXiv:2405.06731 [gr-qc]].


\bibitem{Maeda:2025bnb}
H.~Maeda and C.~Mart\'{\i}nez,
``Exact plane symmetric black bounce with a perfect-fluid exterior obeying a linear equation of state'',
Phys. Rev. D \textbf{112}, no.4, 4 (2025)
doi:10.1103/lt5f-gqmw
[arXiv:2506.14872 [gr-qc]].




\bibitem{Barcelo:2025egc}
C.~Barcel{\'o}, G.~Garc{\'\i}a-Moreno and A.~Jim{\'e}nez Cano,
``Toroidal black holes in four dimensions'',
Class. Quant. Grav. \textbf{42}, no.15, 155015 (2025)
doi:10.1088/1361-6382/adf0e1
[arXiv:2504.16790 [gr-qc]].



\bibitem{Maeda:2024lbq}
H.~Maeda and C.~Mart\'{\i}nez,
``Existence and absence of Killing horizons in static solutions with symmetries'',
Class. Quant. Grav. \textbf{41}, no.24, 245013 (2024)
[addendum: Class. Quant. Grav. \textbf{42}, no.12, 129401 (2025)]
doi:10.1088/1361-6382/ad8ea4
[arXiv:2402.11012 [gr-qc]];




\bibitem{GamboaSaravi}
R.~E.~Gamboa Sarav\'{\i},
``Higher-dimensional perfect fluids and empty singular boundaries'',
Gen. Rel. Grav. \textbf{44}, 1769-1786 (2012)
doi:10.1007/s10714-012-1366-z
[arXiv:1204.4907 [gr-qc]].




\bibitem{Maeda:2018hqu}
H.~Maeda and C.~Mart\'{\i}nez,
``Energy conditions in arbitrary dimensions'',
PTEP \textbf{2020} (2020) no.4, 4
doi:10.1093/ptep/ptaa009
[arXiv:1810.02487 [gr-qc]].


\bibitem{Hawking:1973uf}
S.~W.~Hawking and G.~F.~R.~Ellis,
{\it The large scale structure of space-time},
(Cambridge University Press, Cambridge, 1973).

\bibitem{Iizuka:2025xnd}
N.~Iizuka, A.~Ishibashi, K.~Maeda, H.~Nakayama and T.~Nishioka,
``Energy Conditions and Quantum Information,''
[arXiv:2509.01286 [hep-th]].









\end{references}
\end{document}